\documentclass[twocolumn,aps,showpacs,amsmath,amssymb,preprintnumbers,epsf]{revtex4}
\usepackage{graphicx}
\usepackage{dcolumn}
\usepackage{color}
\usepackage{bm}

\begin{document}

\title{\textbf{Effect of diffusion of elements on network topology and self-organization}}
\author{Ravins and R.K. Brojen Singh}
\affiliation{Centre for Interdisciplinary Research in Basic Sciences, Jamia Millia Islamia,New Delhi 110025, India.}
\date{\today}

\begin{abstract}

{\begin{center}\bf ABSTRACT\end{center}}
{
We study the influence of elements diffusing in and out of a network to the topological changes of the network and characterize it by investigating the behavior of probability of degree distribution ($\Gamma(k)$) with degree $k$. The local memory of the incoming element and its interaction with the elements already present in the network during the growing process significantly affect the network stability which in turn reorganize the network properties. We found that the properties of $\Gamma(k)$ of this network are deviated from scale free type, where the power law behavior contains a exponentially decay factor supporting earlier reported results of Amaral et.al. \cite{ama} and Newman \cite{new1} and recent statistical analysis results on degree distribution data of some scale free network \cite{kha}. Our numerical results also support the behavior of this $\Gamma(k)$. However, we found numerically the contribution from exponential factor to the $\Gamma(k)$ to be very weak as compared to the scale free factor showing that the network as a whole carries the scale free properties approximately.
}
\\\\
\emph KEYWORDS
{: Scale free network, Probability of degree distribution, Master equation, Preferential attachment, Growing network.}
\end{abstract}
\maketitle

Complex networks in which elements interact among each other is recently focused on because these networks fall on mostly in real and natural networks such as scale free and small world networks \cite{str,alb,new,wat,par}, both are microscopic and macroscopic natural networks \cite{alb,sol,per}. Further these networks possess growth and high clustering nodes, which are some of the most important properties of real natural networks, due to activation and deactivation of the nodes in the network \cite{par,per,kle}. If new element is introduced in the network, it may be added or rejected that leads to local perturbation in the network topology which in turn affects in the network stability locally as well as globally \cite{per}. This effect of local stability on the network topology gives rise correlation between network stability and the emergence of self-organization within the network itself \cite{per}. 

The investigation of properties such as self-organization, network stability, signal transduction etc in biological networks in particular has been important area of research in recent times \cite{per}. In these networks, where biochemical reactions are considered to be links among the various participating molecules which are taken as nodes, diffusing in and out of intracellular (various biomolecules) and intercellular molecules (signaling molecules) via intra and extracellular media is an important common phenomena \cite{par}. As a consequence of the diffusing molecules, activation and deactivation of nodes in the network take place that lead to change in the network topology and information transduction within the system and among the systems, where scale free and small world effects (presence of highly clustering nodes) are the signatures \cite{jeo,wag}. 

However, it has been pointed out that many of the real networks do not follow scale free nature since the degree distributions of these networks do not follow power laws but associate with certain exponential component in the distribution function \cite{ama,new1}. It is also argued that if one does proper statistical analysis of the degree distribution data instead of simply fitting the data by straight line in the log-log plot, the degree distribution function follows power law within a certain small range defined by a cut off parameter, $k_c$ and beyond this value it follows a sharp drop-off given by a exponential factor in the distribution \cite{kha}. 

The growth dynamics of a complex network can be described by the change in directed links among the nodes due to the fluctuation in the network topology by incoming and outgoing elements in the network \cite{new1}. Consider $k_i$ be the number of links incident on a particular node $"i"$ or indegree of node $"i"$. Because of this incident node, there should be local perturbation in and around node $"i"$ which in turn affect the network topology. So there could be two possible states of each node : $active$ and $inactive$ states \cite{new1}. The added new node is considered to be always in the active state first and recieves new links from the generated node until it gets deactivated \cite{new1,per}. Once a particular node is deactivated, it will no longer recieve links anymore. Now consider $P(k)$ as the probability of deactivation rate which decreases as in-degree node increases. If we assume this $P$ to be, $P(k)\sim constant+f(k)$, its specific form is considered to be given by the following form,
\begin{eqnarray}
\label{prob}
P(k)\sim \frac{1}{\epsilon}+\frac{\psi}{\eta+k}
\end{eqnarray}
where, $\epsilon$, $\psi$ and $\eta$ are constants provided $\epsilon,\eta\rangle 0$. The first term in the equation (\ref{prob}) is a constant ($\frac{1}{\epsilon}\langle 1$) which indicates that the deactivation rate was initially at constant in the network. The second term incorporates the pertuabation in the deactivation probability when new nodes are added. 

\subsection{Estimation of probability of degree distribution}

Now if we take $\Gamma(k,t)$ to be the in-degree distribution of the active nodes at any instant of time $t$, then the time evolution of $\Gamma(k,t)$ is determined by the following Master equation \cite{new1},
\begin{eqnarray}
\label{master}
\Gamma(k+1,t+1)=\left[1-P(k)\right]\Gamma(k,t)
\end{eqnarray}
Using the equation (\ref{prob}) in (\ref{master}), the following equation can be obtained,
\begin{eqnarray}
\label{master1}
\Gamma(k+1,t+1)-\Gamma(k,t)=-\left[\frac{1}{\epsilon}+\frac{\psi}{\eta+k}\right]\Gamma(k,t)
\end{eqnarray}
Applying boundary condition, $\Gamma(0)=constant$ which means that constant rate of new nodes. The fluctuation is assumed to be small and taken to be constant. Now keeping time to be approximately constant, equation (\ref{master1}) can be written as,
\begin{eqnarray}
\label{master2}
\Gamma(k+1)-\Gamma(k)=-\left[\frac{1}{\epsilon}+\frac{\psi}{\eta+k}\right]\Gamma(k)
\end{eqnarray}
Now, taking $k$ to be continuous, Master equation (\ref{master2}) is reduced to the following differential form,
\begin{eqnarray}
\label{dif}
\frac{\Gamma(k)}{dk}=-\left[\frac{1}{\epsilon}+\frac{\psi}{\eta+k}\right]\Gamma(k,t)
\end{eqnarray}
Solving equation (\ref{dif}), we obtained the expression for $\Gamma(k)$,
\begin{eqnarray}
\label{prob1}
\Gamma(k)=\beta\times e^{-\frac{k}{\epsilon}}(\eta+k)^{-\psi}
\end{eqnarray}
where, $\beta$ is a constant.
\begin{figure}
\label{fig}
\begin{center}
\includegraphics[height=330 pt]{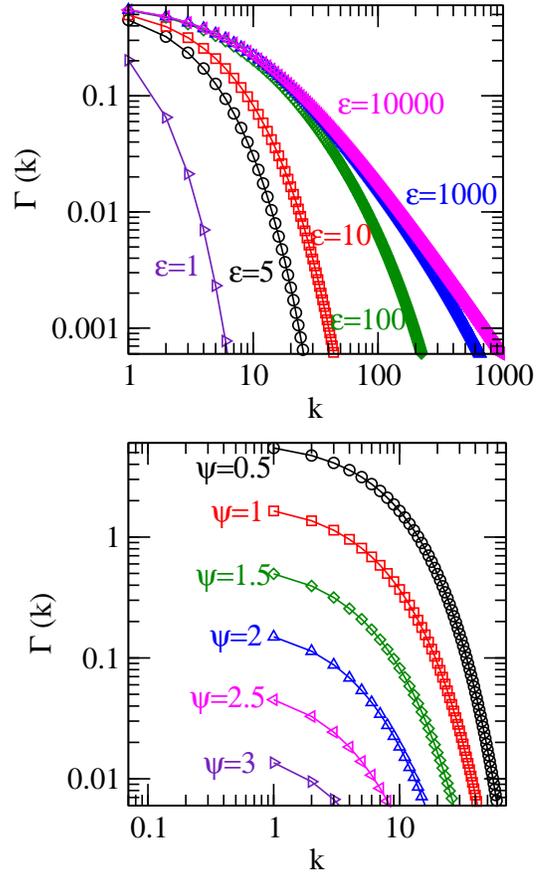}
\caption{Plot of probability of in-degree distribution of active nodes $\Gamma(k)$ as a function of $k$ : (a) $\Gamma(k)$ vs $k$ for different values of $\epsilon$; (b) $\Gamma(k)$ vs $k$ for different values of $\psi$. The values of the constants are: $\eta=10$, $\beta=20$.}
\end{center}
\end{figure}
Since the total number of nodes is large as compared to the number of active nodes, the overall degree distribution, $W(k)$ can be approximated by considering active nodes only, which is given by,
\begin{eqnarray}
\label{deg}
W(k)&=&\frac{d\Gamma(k)}{dk}\nonumber\\
&\sim & \alpha e^{-\frac{k}{\epsilon}\left(\frac{\psi-1}{\psi}\right)}(\eta+k)^{-(\psi+1)}
\end{eqnarray}
where, $\alpha$ is a constant given by, $\alpha=\beta\psi e^{\frac{\epsilon}{\psi\epsilon}}$ and $\psi\epsilon\rangle\rangle (\eta+k)$. The constant $\beta$ in equation (\ref{prob1}) can be calculated by evaluating the normalization condition i.e. $\int_0^\infty \Gamma(k)dk=1$, giving the value, $\beta=(\eta+\frac{1}{\epsilon})^\psi$. Again solving the first moment of the active node distribution, 
\begin{eqnarray}
\label{mom}
m=\int_0^\infty kW(k)dk
\end{eqnarray}
we can estimate the exponent $\psi$. Now defining $\psi=\gamma-1$, where $\gamma$ is the constant related to normalization factor of the probability $\Gamma(k)$ which we obtain after simplification as,
\begin{eqnarray}
\label{exp}
\gamma=1+q(m,\eta,\epsilon)
\end{eqnarray}
Hence, $\gamma$ depends on $m$, $\eta$ and $\epsilon$. 
\begin{figure*}
\label{pvsk}
\begin{center}
\includegraphics[height=650 pt]{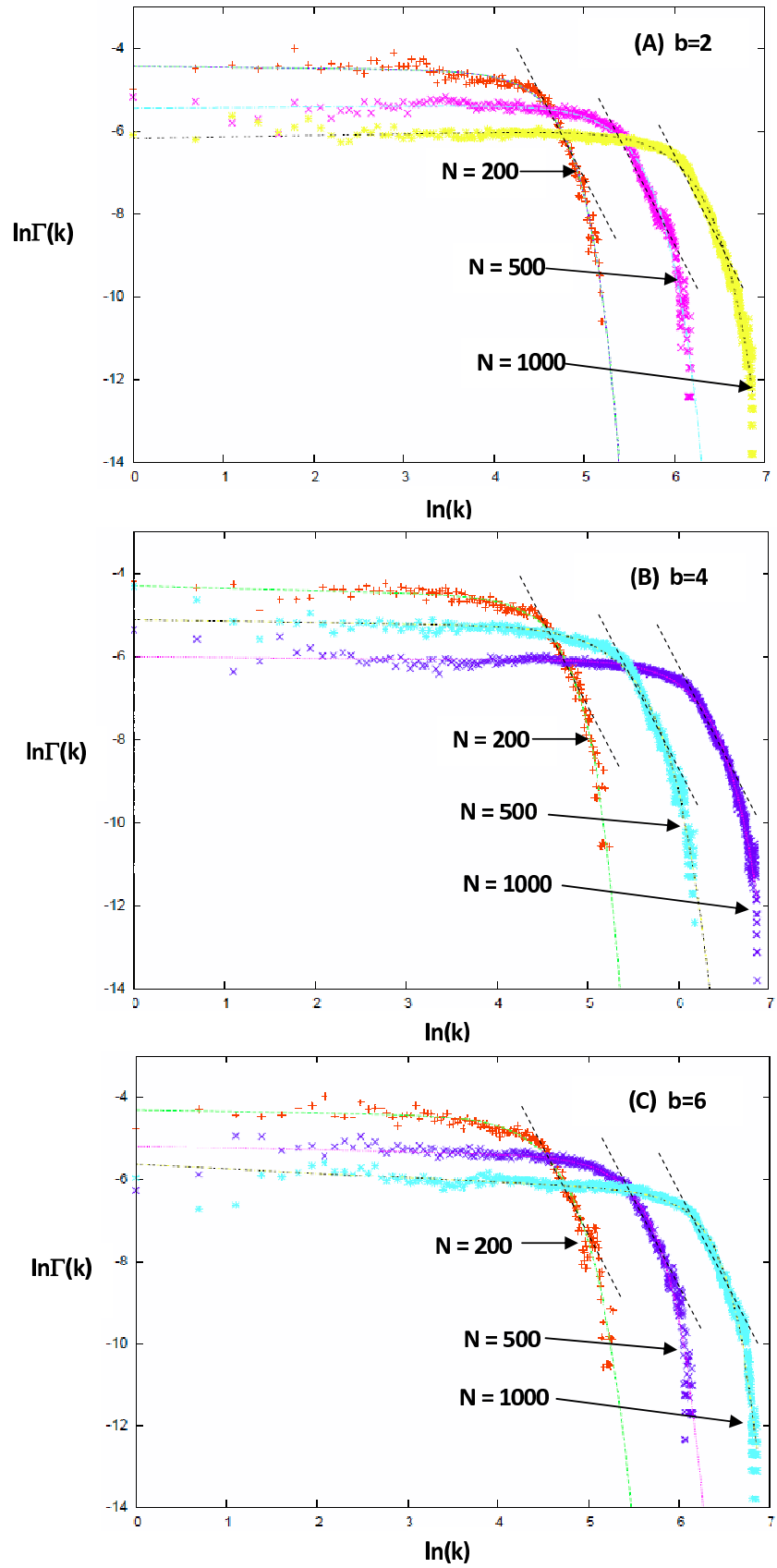}
\caption{Plot of $ln\Gamma(k)$ of active nodes as a function of $ln(k)$ for three different values of $b=2,4,6$, $N=200,500,1000$ and fixed value of $q=3$. The curve lines are fitted curves on the data by $ln\Gamma(k^\prime)=A-Bk^\prime-Ce^{Dk^\prime}$. The dotted straight lines are the fitted lines by straight line $ln\Gamma(k^\prime)=A-Bk^\prime$ to the certain range of data points.}
\end{center}
\end{figure*}

Now, we plot $\Gamma(k)$ in equation (\ref{prob1}) as a function of $k$ for different values of $\epsilon$ and $\psi$ as shown in upper and lower panel of Fig. 1. The plots indicate that for all values of $\epsilon$, the signature of exponential component reflects only in the small values of $k$ with small range. In the remaining large range of $k$, the $\Gamma(k)$ is dominated by power law behavior (straight line nature of $ln\Gamma(k)$-$ln(k)$ data). Moreover, as the value of $\epsilon$ increases, power law part dominate over exponential component in a wider range of $k$.

Now we try to estimate the order parameter $\psi$ from the plot of $\Gamma(k)$ as a function of $k$ for different values of $\psi$ shown in Fig.1 lower panel. The plot shows that $\psi$ is strong for scale free behavior when $\psi\le 2.5$. These characters of $\psi$ and $\epsilon$ indicate that the network of this type has weak contribution in $\Gamma(k)$ from the exponential part, but scale free factor dominates wider range of $k$. This indicates that the overall network properties are dominated and controlled by scale free properties supporting Perotti et al \cite{per}.

\subsection{Preferential attachment}

If $\Lambda(k)$ is the degree dependent attachment rate and the network contains $t$ nodes, then $tW(k)$ of these will have degree $k$. The number of active nodes with degree $k$ is $m\Gamma(k)$. Hence, the average increase in degree is given by, $\Lambda(k)=\frac{m\Gamma(k)}{tW(k)}$. Using equation (\ref{prob1}), we get the preference attachment rate as,
\begin{eqnarray}
\label{pre}
\Lambda(k)=Z(\eta+k)e^{-\frac{k}{\epsilon\psi}}
\end{eqnarray}
where, the constant $Z$ is given by, $Z=\frac{mc}{t\alpha}$.

\subsection{Numerical results}

We follow the numerical algorithm due to Perotti et al \cite{per} to calculate probability of degree distribution, $\Gamma(k)$ of growing network generated by diffusion of nodes in and out of the network. The algorithm is based on generated memory of grow or shrink at every step characterized by eigen value ($\lambda$) of the modified Jacobian matrix of the elements in the network at that step. If $\lambda~\langle~ 0$ then the new node diffused inside the network is accepted by introducing a range of interaction range defined by a uniform random number $b$ in $[-b,b]$. However the condition $\lambda~\rangle~0$ destablize the network, a condition is imposed that either the node is eliminated or certain number of already existed nodes ($q\le k_{n+1}$) in the previous step are eliminated.  
\begin{figure}
\label{pr}
\begin{center}
\includegraphics[height=180 pt]{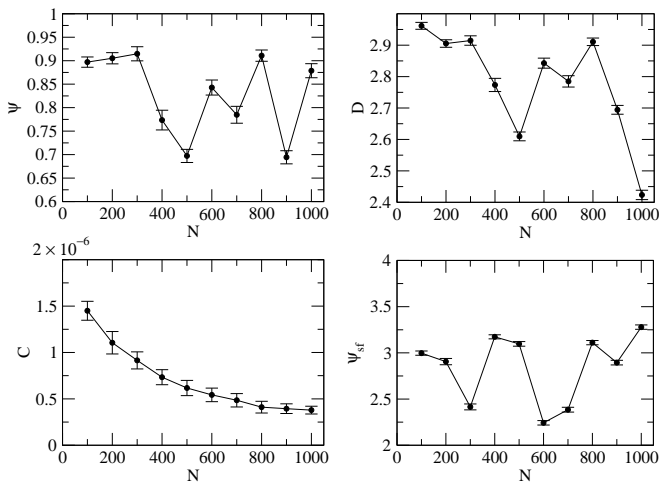}
\caption{Plots of different parameters as a function of $N$ : (a) $\psi$ versus $N$, (b) $D$ versus $N$, (c) $C$ versus $N$ and (d) $\psi_{sf}$ versus $N$. The error bars are coming from the calculation of different parameters from various data sets i.e. various data of $b$, $q$ and $N$.}
\end{center}
\end{figure}

Now we present our numerical results based on the algorithm we explained for different values of $b=1,2,4,6,8,10$ and $q=2,3,4$ as shown in Fig.2. We calculated $\Gamma(k)$ as a function of $k$ for the mentioned parameters and $log\Gamma(k)-log(k)$ for $b=2,4,6$ for $q=3$ is shown in the plots. The the curve lines in each plot are the fitted curves using our simplified expression of $\Gamma(k)$ in equation (\ref{prob1}) (we take $\eta=0$ for the sake of simplicity)i.e. $ln\Gamma(k^\prime)=A-Bk^\prime-Ce^{Dk^\prime}$, where, $k^\prime=ln(k)$ and $A$, $B$, $C$ and $D$ are constants to be obtained by fitting the curves. The equation fits well the data but the order parameter $\psi$ is found to be very small as compared to the predicted values of scale free networks. The parameter $\psi$ is calculated as a function of network size ($N$), by averaging its values calculated for various values of parameters $b$ and $q$, and is shown in Fig. 3 (left upper panel). Taking mean of the values for different values of $b$, $q$ and $N$, we found $\psi=0.83\pm 0.053$. Then we fit the curves with straight lines so that we can get the values of scale free parameter $\psi_{sf}$ for the curves (fitted straight lines are shown in Fig.2) as a function of $N$. The result is shown in Fig.3 (right lower panel) and the average value of $\psi_{sf}$ is found to be $2.78\pm 0.127$.

We then search for the range of $k$ within which scale free properties work by looking at the values of exponential factor in the equation (\ref{prob1}). So we obtain the values of $C$ from the curve fitting procedure in all cases and is found to be very small and decreasing as $N$ increases and become approximately stationary after $N\sim 800$ ($C=0.47\times 10^{-6}$). The plot of $C$ versus $N$ is shown in lower left panel of Fig.3. The scale free works in a small regime of $k$ for all network sizes $N$, for example for $N=1000$, it works within the regime $412~\langle~k~\langle~894$ approximately. 

We now search for the values of $D$ as a function of $N$ following the fitting process of the same set of data and is shown in Fig.3 upper right panel. The average value of $D$ is found to be $2.6\pm0.256$. Putting all these values of constants, we can write approximately the expression for $\Gamma(k)$ as in the following,
\begin{eqnarray}
\label{num}
\Gamma(k)\approx constant\times e^{-4.7\times 10^{-7}k^{2.6}}\times (\eta+k)^{-\psi}
\end{eqnarray}
This means that the exponential factor influence the scale free properties slightly. The equation (\ref{num}) further shows that even if the scale free nature works in a small range of $k$, it dominates and carries the overall network properties. Eventually, the order parameter should scale by $\psi\rightarrow\psi_{sf}$ when one considers overall network topology.

The activation and deactivation of nodes in a network due to diffusing of elements in and out of the network, provided the probability of deactivation rate we have taken in equation (\ref{prob}), definitely give a change in the in-degree distribution of active nodes. This distribution contain a power in $k$ factor with an exponential component that makes a different distribution from scale free type. When $\epsilon\rangle\rangle k$, $\Gamma(k)$ follows power law falling to scale free behavior. But when $\epsilon$ is comparable to $k$ value, $\Gamma(k)$ is deviated from scale free type. However the overall network carries the scale free properties as evident from our analytical results.

Since the exponential factor contributes a little in the $\Gamma(k)$ as evident from our numerical results, the overall network properties fall into scale free type. This indicates that the perturbation due to diffusion of nodes in and out of the network influnce the properties of network structure as indicated by the deviated value of order parameter $\psi$. However still the scale free properties dominate the overall network properties indicating the self-organization of the network after the pertubation is induced.

{\bf Acknowledgments}
This work is financially supported by University Grants Commission (UGC) and carried out in center for Interdesciplinary research in basic sciences, Jamia Millia Islamia,New Delhi,India.

\end{document}